# Quantum size phenomena in single-crystalline bismuth nanostructures

*Egor A. Sedov[1], Kari-Pekka Riikonen[2] and Konstantin Yu. Arutyunov[1,3]\**

[1]National Research University Higher School of Economics, Moscow Institute of Electronics and Mathematics,109028, Moscow, Russia

[2]Nano Science Center, Department of Physics, University of Jyväskylä, PB 35, 40014, Jyväskylä, Finland

[3]P.L. Kapitza Institute for Physical Problems RAS, 119334, Moscow, Russia

\* karutyunov@hse.ru

## Abstract

Size-dependent quantization of energy spectrum of conducting electrons in solids leads to oscillating dependence of electronic properties on corresponding dimension(s). In conventional metals with typical energy Fermi $E_F \sim 1$ eV and the charge carrier's effective masses m\* of the order of free electron mass $m_0$, the quantum size phenomena provide noticeable impact only at nanometer scales. Here we experimentally demonstrate that in single-crystalline semimetal bismuth nanostructures the electronic conductivity non-monotonously decreases with reduction of the effective diameter. In samples grown along the particular crystallographic orientation the electronic conductivity abruptly increases at scales of about 50 nm due to metal-to-insulator transition mediated by the quantum confinement effect. The experimental findings are in reasonable agreement with theory predictions. The quantum-size phenomena should be taken into consideration to optimize operation of the next generation of ultra-small quantum nanoelectronic circuits.

# Introdution

With reduction of dimension(s) classic physics gradually degenerates into quantum. For example, energy separation between the *n*-th and the (*n+1*)-th energy levels of a particle with mass *m* confined to a space of characteristic dimension *a* is $\Delta E(n) \sim n/ma^2$ given $n \gg 1$. This key quantum feature is in the origin of discrete energy spectra of atoms. In metals size quantization affects basically all electronic properties: conductivity, Hall and thermoelectric coefficients, reflectivity, chemical surface activity, etc.[1] For a typical metal with effective mass $m^* \approx m_0$ at realistic temperatures $k_B T < \Delta E$ one should expect quantum size phenomena to manifest at difficult-to-achieve scales $a \sim 1$ nm. A solution is to study materials with $m^* \ll m_0$ such as bismuth, antimony or *BiSb* alloys [2]. Peculiarity of the energy spectrum of bismuth is the co-existence of two types of charges: 'light' L-electrons with $m_e^* \ll m_0$ and 'heavy' T-holes with $m_h^* \sim m_0$ (Fig. 1a,b). The heavy holes 'stabilize' Fermi energy $E_F$ leading to oscillating dependence of transport properties vs. characteristic dimension, and pronounced decrease of electron resistivity due to metal-to-insulator transition at the threshold scale $a_0$ so that $E_F \sim 1/m_e^* a_0^2$. As bismuth energy spectrum is highly anisotropic (Fig. 1a), the size-dependent behavior of characteristic parameter, e.g. electric conductivity, depends on particular crystallographic orientation.

At early stages of experimental studies [3,4] 2D films were of primary attention, while later other low-dimensional bismuth structures have been investigated: nanowires [5,6], microcylinders [7] and point contacts [8,9]. Experiments with wide polycrystalline films [3,4] or bundles of nanowires [6] can be interpreted only qualitatively as the contributions of individual grains / nanowires are averaged. STM experiments [8,9] have demonstrated clear size-dependent variation of electric conductivity. However comparison of STM data with theory is ambiguous as neither the crystallographic orientation, nor the shape of an STM-pulled point contact are clearly defined.

Here we present experimental study of electric conductivity of single-crystalline bismuth nanostructures (or alternatively they can be called *nanorods*, given their relatively short length compared to width and thickness). The oscillatory dependence of resistance on sample cross section is in good agreement with model predictions. The pronounced increase of resistivity below a certain scale can be interpreted as manifestation of size-mediated metal-to-insulator transition.

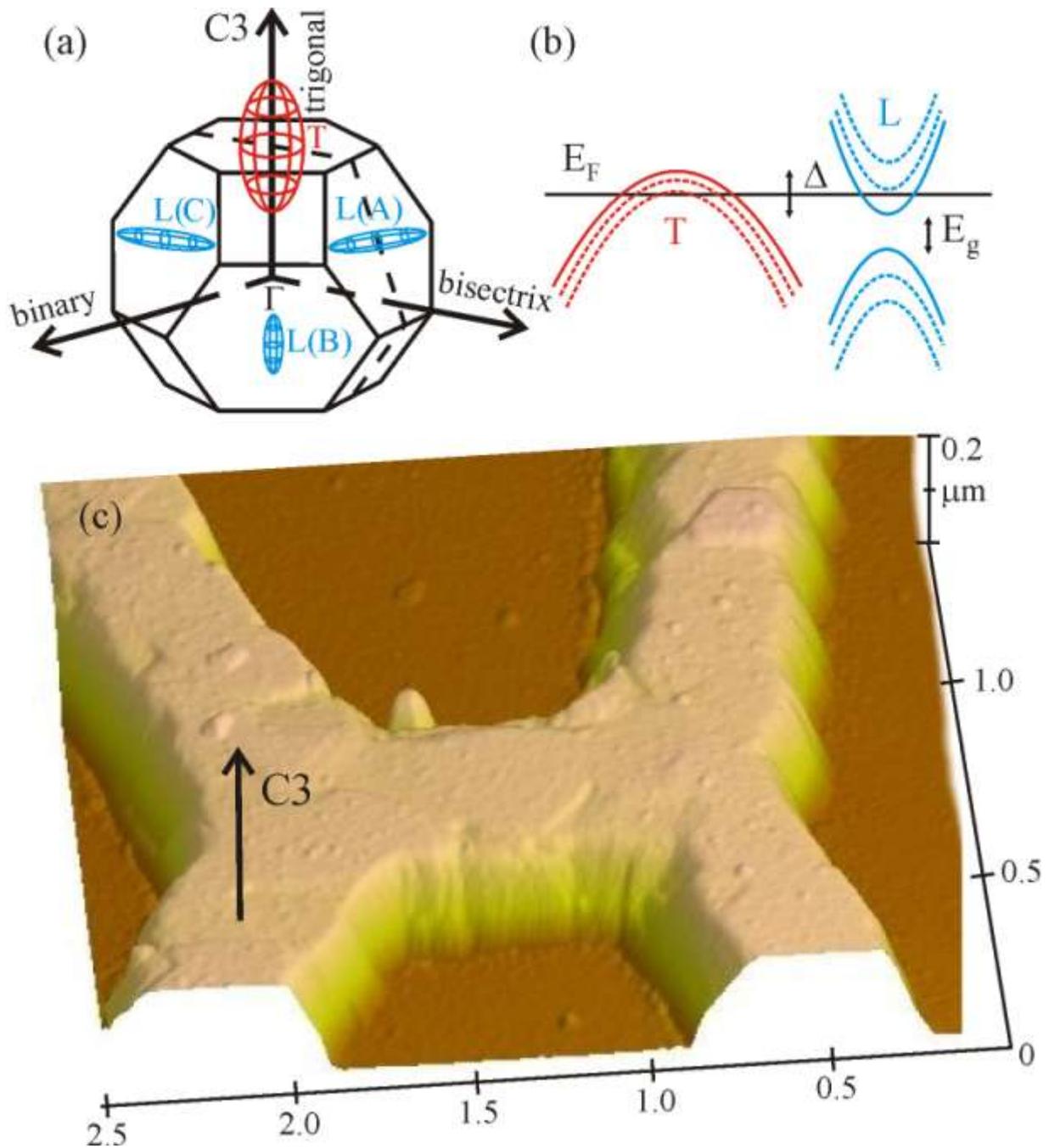

**Fig. 1. Fermi surface, energy spectrum and typical nanostructure.** (a) Fermi surface of bulk bismuth. Electron 'pockets' are presented by ellipsoids in three equivalent L-points of the Brillouin zone, holes are located in ellipsoid in T-point. (b) Schematic representation of the energy spectrum of bismuth. Solid lines correspond to bulk sample, while the dashed lines stand for energy quantization due to quantum size effect. (c) AFM image of typical structure before ion milling. One can clearly distinguish the single crystalline texture. Trigonal axis C3 is normal to the sample plane.

**Theory background**

There have been numerous approaches of various complexity to describe energy spectrum of bulk bismuth[10]. For our purposes it is quite sufficient to consider rather simple model [11,12]. For T-point holes (Fig.1 a,b) the energy $E$ vs. momentum $\hbar \mathbf{k}$ dependence is:

$$E^h(k) = \frac{\hbar^2 k_x^2}{2m_x^h} + \frac{\hbar^2 k_y^2}{2m_y^h} + \frac{\hbar^2 k_z^2}{2m_z^h} \quad (1),$$

where the diagonal components of hole effective mass tensor (in units of free electron mass $m_0$) are $m_x^h = m_y^h = 0.059$, $m_z^h = 0.634$. For L-point electrons ("+") and L-point holes ("-"):

$$E^e(k) = -\frac{E_g}{2} \pm \frac{E_g}{2}\sqrt{1 + \frac{2\hbar^2}{E_g}\left(\frac{k_x^2}{m_x^e} + \frac{k_y^2}{m_y^e} + \frac{k_z^2}{m_z^e}\right)} \quad (2),$$

where the corresponding masses along the bisectrix axis are $m_x^e = 0.00139$, $m_y^e = 0.291$, $m_z^e = 0.0071$. There is the consensus between various literature sources that the energy gap in L-point is $E_g = 13.7 \pm 0.1$ meV, and the overlapping between T and L-bands is $\Delta = 38$ meV (Fig. 1b).

Quantum size phenomena in bismuth have been considered already at early ears of 'bismuthology' [2,13,14]. For interpretation of our data we utilize rather straightforward model based on Boltzmann equation [15]. Following Ref. 15 and considering bismuth nanowire of rectangular cross section $w \cdot t$, where $w$ is width and $t$ is thickness, one gets the oscillatory behavior of electronic conductivity on corresponding dimensions:

$$\sigma^e = \frac{2e^2}{\pi \hbar S} \frac{\mu_x^e}{m_z^e} \sum_{ij}^{[r_w][r_t]} \frac{(2\hbar/V_0)^2 \sqrt{U_{ij}^e}}{\sum_{i'j'}^{[r_w][r_t]} \Lambda_{i'j'}^{ij} \sqrt{U_{i'j'}^e}} \quad (3),$$

where $S$ is the volume concentration of scatterers each of strength $V_0$, $e$ is electron charge; outer summation stands for contribution of different sub-bands $(i,j)$; $\Lambda^{mn}_{m'n'} \equiv (2+\delta_{mm'})(2+\delta_{nn'})$ with $\delta_{ab}$ being Kronecker delta function; $U^e_{ij} \equiv 1-(i/r_w)^2 + (\mu^e_y/\mu^e_x)[1-(j/r_t)^2]$; $[z]$ stands for integer part of argument $z$. Parameters $r_w \equiv w/w_0$ and $r_t \equiv t/t_0$ are width and thickness of nanowire, normalized by the relevant dimensions $w_0=(h/2)(M_x\Delta_x)^{1/2}$ and $t_0=(h/2)(M_y\Delta_y)^{1/2}$ corresponding to metal-to-insulator transition: when gap opens between the lowest L-electron and the top-most T-hole bands (Fig. 1b). Here $M_{x,y} \equiv (m^e_{x,y} m^h_{x,y})/(m^e_{x,y} + m^h_{x,y})$ and $\mu^e_{x,y} \equiv (\Delta_{x,y} m^h_{x,y})/(m^e_{x,y} + m^h_{x,y})^{-1}$, $\Delta_{x,y}$ is the energy bands overlap due to confinement of charge carriers in corresponding direction $x$ or $y$. Parameters $S$ and $V_0$ are not known with necessary accuracy and are difficult to be determined from transport experiments. Contributing to absolute value of electron conductivity, they can be considered as adjustable parameters to fit $\sigma^e(w,t)$ data for thick wires $w \to \infty$ and $t \to \infty$ when size phenomena can be neglected. Similar to Eq. 3 expression for hole conductivity can be obtained and should be taken into consideration to calculate the total electric conductivity. Given much larger values of T-hole masses compared to L-electrons along bisectrix axis (Fig. 1a), we consider the hole conductivity $\sigma^h(w,t)$ as constant and neglect the size-dependent component.

The $\sigma^e(w,t)$ characteristic depends on particular orientation of the sample against crystallographic axes. Simple $\sigma^e(w,t)$ dependence appears for high-symmetric directions, while along an arbitrary direction the corresponding characteristics might be rather complicated containing secondary maxima originating from contributions of all non-equivalent pockets of bismuth Fermi surface (Fig. 1a). Note that even for the most favorable orientation along bisectrix axis C2 corresponding to the lowest electron masses $m^e_x$, the actual $\sigma^e(w,t)$ dependence is affected not only by the cross section ($w \cdot t$), but also by the actual values of width $w$ and thickness $t$ (Fig. 2).

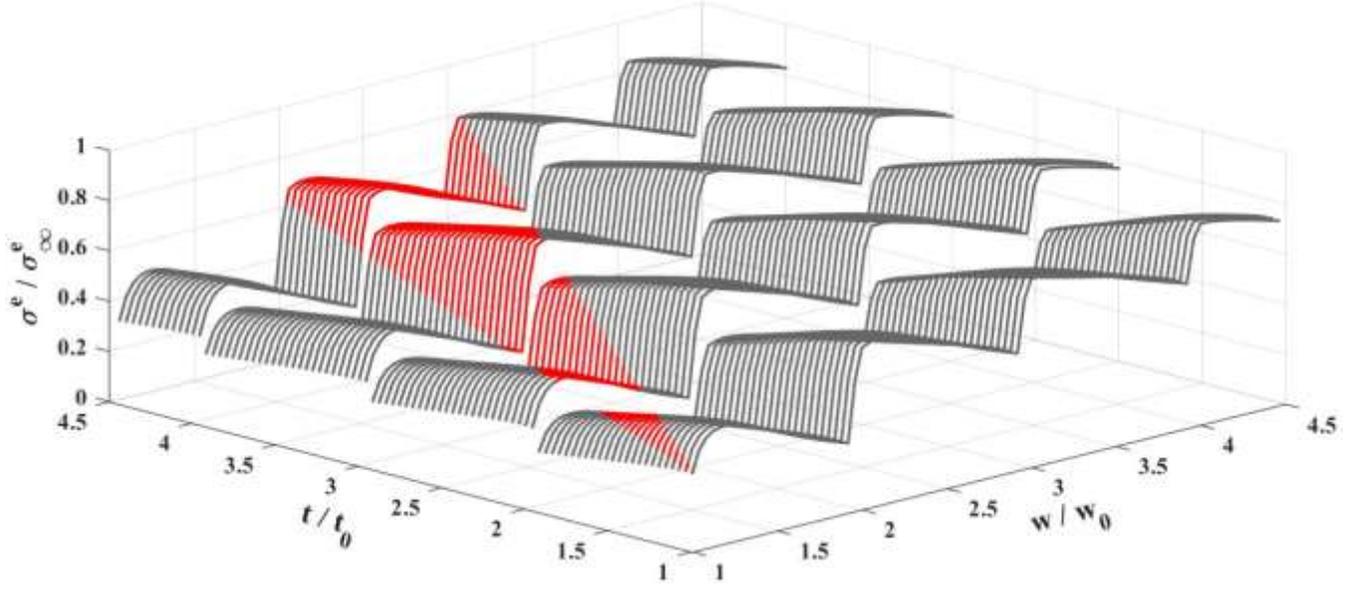

**Fig. 2. Calculated electronic conductivity of bismuth nanostructure cut along bisectrix axis based on model [15]**. For nanowire of rectangular cross section (*wt*), where *w* is width and *t* is thickness, the electronic conductivity $\sigma^e$ exhibits oscillatory behavior as function of corresponding dimension(s) in accordance with Eq. 3. The bulk conductivity $\sigma^e_\infty$ recovers at scales $w \gg w_0$, $t \gg t_0$, where $w_0 \approx 110$ nm and $t_0 \approx 25$ nm are the critical width and thickness corresponding to estimated metal-to-insulator transition (for details see comments below Eq. 3). Bright cone represents the range of sample dimensions studied in present work with actual cross sections, obtained by AFM analyses, equal to cross sections (*wt*) of an effective rectangular nanowire [15].

## Methods

Experimental confirmation of a size phenomenon typically involves studies of multiple structures with various values of the relevant dimension: e.g. thickness of a film or diameter of a wire. The approach assumes that all other parameters of the system, except the dimension(s), are the same. Unfortunately in practice the requirement it is not easy to satisfy, particularly if one deals with various artificially fabricated state-of-the-art small nanostructures. An alternative approach is to repeatedly re-measure the same sample after successive steps of size reduction [16,17]. Low energy directional ion milling [18,19] was used to progressively and non-destructively reduce the diameter of bismuth nanostructures. The penetration depth of Ar$^+$ ions accelerated to 1 keV into bismuth matrix is about 1 nm, which is comparable to thickness of just few top-most atomic layers. Hence, the method can be considered as virtually non-destructive[17]. Extra information on ion milling is provided in *Supplementary information* file. AFM and SEM analyses were made *ex situ* at room temperatures and atmosphere after each sputtering and resistance measurement sequence. As bismuth is a noble material, contamination of the surface due oxidation is not an issue.

Quantitative analysis of quantum size effects in a material with highly anisotropic Fermi surface as bismuth (Fig. 1a) requires fabrication of single-crystalline nanostructure with well-defined orientation of crystallographic axes. Several approaches have been taken to fabricate single crystalline micro- or nanowires of various materials [6,7,20,21]. Here we used conventional e-beam lift-off lithography and high speed ~1 nm/s 99.9995% pure bismuth e-gun deposition in vacuum ~ $10^{-6}$ mBar on mica substrate heated to 140 °C. The recipe provides high-quality polycrystalline films with large grain size up to 1.5 µm. X-ray analyses of co-deposited 2D films revealed that trigonal axis C3 is normal to the film plane (Fig. 1c), which is in agreement with literature data [3,4,14]. Combined with ion beam milling [18,19] the method enables sequential electron transport measurements of bismuth nanostructures with progressively reduced transverse dimension $\sim(t \cdot w)^{1/2}$ down to sub-50 nm scales. All structures were designed to enable electric resistance measurements using 4-probe configuration to reduce the impact of the probes (Fig. 1c). The absence of grain boundaries containing foreign material(s) (e.g. impurities or oxide) is confirmed by the linearity of V-I dependencies, typical for Ohmic conductors.

## Results and discussion

The experiment consisted of bismuth thin films and nanostructures resistance *R* measurements as function of transverse dimensions *w* and *t* between sessions of ion milling. All samples, including the co-deposited and co-sputtered 2D films, demonstrated non-monotonous $R(t,w)$ dependencies (Fig. 3). Note that electron transport data obtained on wide granular 2D films (Fig. 3a), similar to early studies [2,3,4], can be interpreted only qualitatively as orientation of electric current vs. crystallographic axes in each grain is different. For sufficiently long nanowires with length *l* noticeably exceeding the characteristic grain size, as studied in Ref. 21, an agreement between experiment and theory is speculative. The problem originates from random impact of each of the *i*-th grain connected in series. Superposition of all those contributions results in complicated overall $R = \sum_i R_i(d_x^i, d_y^i, d_z^i)$ dependence, difficult to analyze as the orientation of each grain and its dimensions {$d_x^i$, $d_y^i$, $d_z^i$} are not known.

Here we focus on quasi-1D structures $w,t \ll l \sim 1$ μm with the 'body' formed from a single grain (Fig. 1c), where the $R(t,w)$ dependencies can be modeled with reasonable accuracy. For the most favorable orientation of the sample along bisectrix axis (Fig. 1a), the dominating impact comes from L-electrons with minimal electron effective mass $m_x^e \approx 0.0014 m_0$. In this geometry the size-dependent contributions from two other electronic L-pockets and T-holes with much larger effective masses are noticeably weaker. Certainly, the absolute value of total electric conductivity is the sum of contributions from all charge carriers.

$R(t,w)$ data with corresponding theory dependence is presented in Fig. 3b. The experiment shows both the non-monotonous variation of sample resistance vs. transverse dimensions $R(t,w)$, and the pronounced increase of resistance below the threshold $(t_0 \cdot w_0)^{1/2} \sim 50$ nm. The non-monotonous variation of resistance $R(t,w)$ originates from the size-dependent quantization of the energy spectrum (Fig. 1b), with minima of resistance (maxima of conductivity) corresponding to matching of the Fermi energy $E_F$ and the $n$-th energy level. The increase of the resistance below 50 nm is related to opening of the gap in the energy spectrum: the size-dependent transition from metallic to dielectric state. Yet another manifestation of the same effect has been recently observed as non-monotonous dependencies of power factor and thermoelectric figure of merit in Bi nanowires at sub-300 nm scales [22]. The agreement between experiment and theory is acceptable, but not perfect: the main features do match, while the 'fine structure' predicted by theory is not observed. Presumably several factors contribute to the discrepancy. First, finite temperature $T=4.2$ K corresponds to broadening of each size-quantized energy level by ~0.5 meV, leading to smearing of features predicted by the essentially zero-temperature model [15]. The observation is supported by data in Fig. 3, which indeed show much weaker $R(t,w)$ dependence at room temperature compared to $T=4.2$ K. Second, due to requirement to fabricate single-crystalline samples of minimal size suitable to be ion milled to reduce transverse dimensions, the measuring probes are of comparable dimensions as the 'body' of the sample (Fig. 1c). Hence, the electrodes cannot be considered as non-invasive and size quantization phenomena inside the probes might provide non-negligible contribution to measured signal. Third is the uncertainty in orientation of crystallographic axes with respect to direction of bias current ($x$). For data presented in Fig. 3b the misorientation between the bisectrix axis C2 and the sample axis $x$, deduced from theory fits with Eq. 3, is just about few degrees. The effect of such a small misorientation was formally taken into account as small adjustment (increase) of the effective mass $m_x^e$ in Eq. 2. Note that the model from Ref. 15 assumes all current lines being parallel to each other. That would be the case if the length of the sample $l$ is much larger than the transverse dimensions $w$ and $t$, which is not the case particularly for 'thick' samples before ion milling (Fig. 1c). Better agreement with experiment formally can be achieved assuming size-dependent contribution of more than one crystallite (e.g. from nearby regions of the probes) to the measured integral signal. However, as those contributions are not clearly defined, such 'improvement' of fits would be speculative.

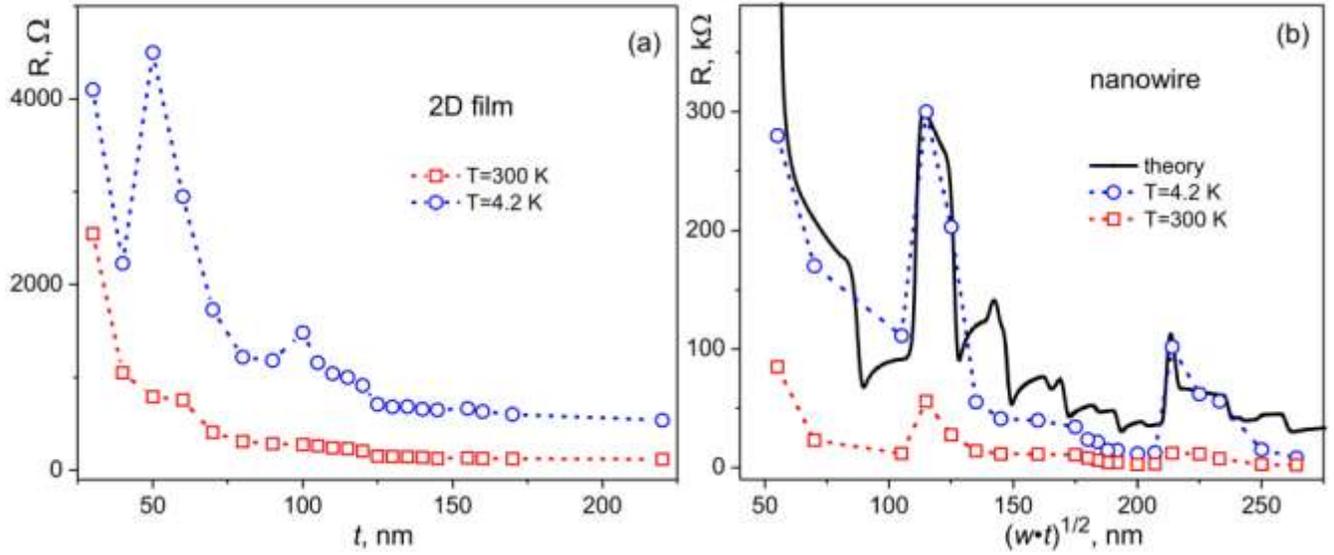

**Fig. 3. Size dependence of resistance.** (a) Typical dependence of co-deposited wide 2D bismuth film resistance $R$ on thickness $t$. (b) Dependence of bismuth nanowire resistance $R$ on effective diameter $d_{eff}=(w \cdot t)^{1/2}$. Circles (○) correspond to liquid helium temperature T=4.2 K, squares (□) stand for room temperature data. Dotted lines connecting experimental points are guides for eye. Theory fit [15] (solid line) assumes trigonal axis C3 being perpendicular to the sample plane and ~3 degree misorientation angle between the sample axis and the crystallographic bisectrix axis C2. Best fit 'trajectory' in coordinates $R(w,t)$ (see Fig. 2) stands for reduction of δw=1.87 nm and δt=1.69 nm between successive points starting from a nanowire with initial width w=300 nm and thickness t=265 nm.

To conclude, we fabricated and measured resistance of single-crystalline bismuth nanostructures as function of transverse dimensions: thickness $t$ and width $w$. The $R(t,w)$ data showed clear non-monotonous dependence with pronounced increase of resistance approaching the threshold diameter ~

50 nm. Both effects are more pronounced at low temperatures. Our findings are consistent with model calculations taking into consideration size-dependent quantization of the energy spectrum of charge carriers in bismuth. The observed quantum size effect is of universal origin and should be present in any metallic conductor of relevant dimensions, and, hence, should be taken into consideration to optimize operation of the next generation of ultra-small quantum nanoelectronic circuits.

**Acknowledgements**

The article was prepared within the framework of the Academic Fund Program at the National Research University Higher School of Economics (HSE) in 2016 (grant No. 16-05-0029) and supported within the framework of a subsidy granted to the HSE by the Government of the Russian Federation for the implementation of the Global Competitiveness Program

**Author contributions**

Egor A. Sedov (MSc student) made calculations and theory fits. Kari-Pekka Riikonen (MSc student) fabricated nanostructures, made microscopic analyses and performed transport measurements. Konstantin Yu. Arutyunov (Prof. ) initiated the project, designed variable temperature sample holder in the vacuum chamber, built measuring set-up, participated in transport measurements, suggested interpretation of the results and wrote the text.

**Competing financial interests**

Authors do not have any competing financial interests

# Supplementary information

*Quantum size phenomena in single-crystalline bismuth nanostructures*

*Egor A. Sedov, Kari-Pekka Riikonen and Konstantin Yu. Arutyunov*

Here we provide extra information about ion milling of Bi nanowires.

Each sample was analyzed using scanning electron SEM and atomic force AFM microscopes. Examples of AFM images of the same nanostructure just after fabrication and after several sessions of ion milling are presented in *Supplementary-Fig1*. One can clearly distinguish crystalline structure of the original structure (left panel), which is somehow less pronounced after ion treatment (right panel). The 'halo' (or 'terrace') along the perimeter of the processed sample corresponds to sputtered mica substrate.

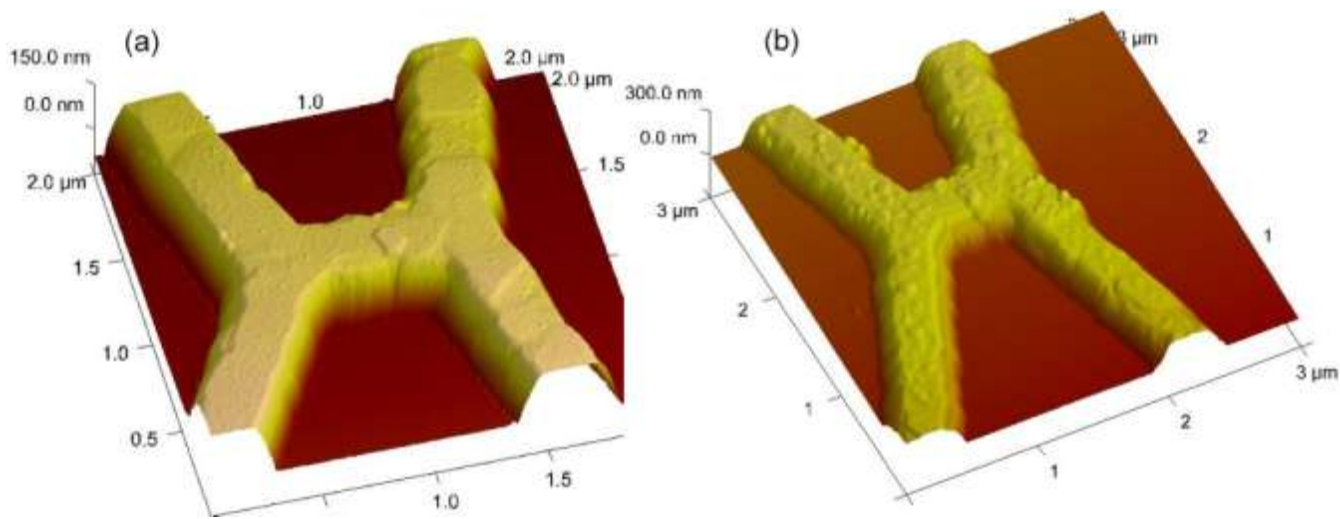

**Supplementary-Fig1**. AFM images of the same bismuth nanostructure just after fabrication (a) and after several sessions of ion milling (b). Note the 'rounding' of all edges and imperfections due to sputtering.

There are two main sources of uncertainty in determination of sample cross section using AFM. First, applicable both for original and ion milled structures, originates from attraction forces acting on AFM tip not only in Z-direction, but also in XY plane when the tip approaches edges of an essentially 3D structure. Typically this effect leads to overestimation of nanostructure cross section. Second uncertainty is important only for ion milled samples. It is quite problematic to distinguish the boundary between the sputtered metal and the substrate. Making multiple calibrations and test measurements defining the rate of ion milling for bismuth and mica we can estimate the position of the boundary with accuracy ± 3 nm. Combined error in determination of a nanostructure cross section might reach 40% for multiple times ion milled (thinnest) samples (*Supplementary-Fig2*). Detailed analyses of the sputtering process and the related errors are provided in Ref. 19.

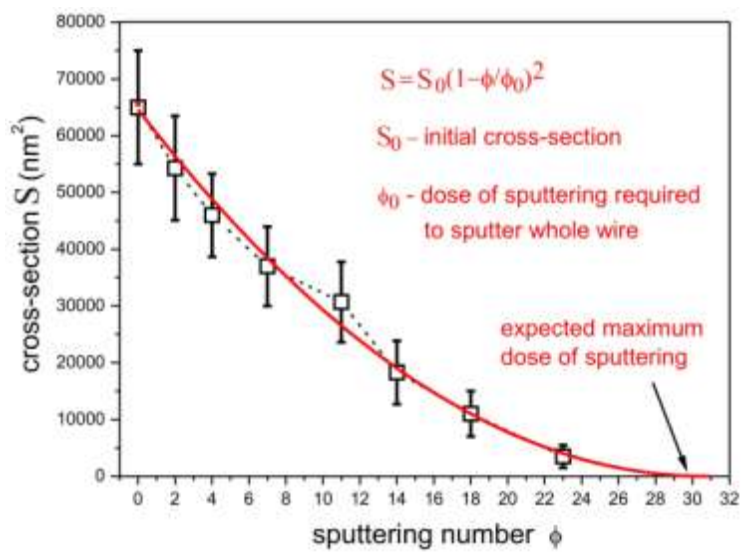

**Supplementary-Fig2**. Sample cross section vs. sputtering. Open squares (□) and corresponding error bars stand for AFM data. Dashed line is guide for eyes. Solid line is the parabolic approximation of the trend (for details see Ref. 19).